# Investigation of Magnesium Silicate as an Effective Gate Dielectric for AlGaN/GaN Metal Oxide High Electron Mobility Transistors (MOSHEMT)

Seshasainadh Pudi[1], Navneet Bhardwaj[1], Ritam Sarkar[4], V S Santhosh N Varma Bellamkonda[1], Umang Singh[1], Anshul Jain[3], Swagata Bhunia[2], Soumyadip Chatterjee[1], and Apurba Laha[1]

[1]Department of Electrical Engineering, Indian Institute of Technology Bombay, Mumbai 400076, India
[2]Department of Physics, Indian Institute of Technology Bombay, Mumbai 400076, India
[3]Center for Research in Nano Technology and Sciences, Indian Institute of Technology Bombay, Mumbai 400076, India
[4]Department of Electrical Engineering, Indian Institute of Technology Bombay, Mumbai, India, (*Current affiliation- IMEC, Leuven, Belgium)

Email: seshasainadh2020@gmail.com, ritam.sarkar@imec.be, laha@ee.iitb.ac.in

*Abstract* — In this study, a 6 nm layer of Magnesium Silicate (Mg-Silicate) was deposited on AlGaN/GaN heterostructure by sputtering of multiple stacks of MgO and $SiO_2$, followed by rapid thermal annealing in a nitrogen ($N_2$) environment. The X-ray photoelectron spectroscopy (XPS) analysis confirmed the stoichiometric Mg-Silicate (MgSiO3) after being annealed at a temperature of 850°C for 70 seconds. Atomic force microscopy (AFM) was employed to measure the root mean square (RMS) roughness (2.20 nm) of the Mg-Silicate. A significant reduction in reverse leakage current, by a factor of three orders of magnitude, was noted for the Mg-Silicate/AlGaN/GaN metal–oxide–semiconductor (MOS) diode in comparison to the Schottky diode. The dielectric constant of Mg-Silicate ($\varepsilon_{Mg\text{-}Silicate}$) and the interface density of states ($D_{it}$) with AlGaN were approximated at ~6.6 and $2.0 \times 10^{13}$ $cm^{-2}eV^{-1}$ respectively, utilizing capacitance-voltage (CV) characteristics.

# 1. INTRODUCTION

The field of high frequency and power electronics has significantly advanced over the past few decades, thanks to the development of GaN-based high electron mobility transistors (HEMTs). The unique features of GaN material, such as a high breakdown field, high saturation velocity, high thermal conductivity, high electron mobility, and the formation of a two-dimensional electron gas density (2-DEG), make it a superior candidate for high frequency and high-power devices [1]. Consequently, AlGaN/GaN HEMTs have become the workhorse of high-power, high frequency, and power switching applications. Nevertheless, gate leakage issues in conventional AlGaN/GaN HEMTs tend to lower their performance considerably. To mitigate this, metal-insulator-semiconductor HEMTs (MIS-HEMTs) have been proposed and demonstrated by various research groups. Essential attributes of the insulator used for MIS-HEMTs include a high dielectric constant, a smooth interface with the AlGaN surface, and a high band gap. Numerous oxides and insulators have been explored as gate dielectrics for MIS-HEMTs, including $SiO_2$ ($\varepsilon_{SiO2}$ =3.9) [2-3], $Al_2O_3$ ($\varepsilon_{Al2O3}$ =10) [6-10], $HfO_2$ ($\varepsilon_{HfO2}$ = 20) [11-14] $ZrO_2$ ($\varepsilon_{ZrO2}$=23) [15-16], $Ta_2O_5$ ($\varepsilon_{Ta2O5}$ =11.8) [17-18], $TiO_2$ ($\varepsilon_{TiO2}$ = 25) [19-22], $Ga_2O_3$ [23], $Gd_2O_3$ [24-26], $Sc_2O_3$ [27], $Nb_2O_5$ [28-29], and $Si_3N_4$ [30]. The silicate is an insulator material, which could be a suitable candidate for a gate dielectric. Magnesium silicate has high dielectric constant (~6.6) and wide band gap, due to which it has the potential to use a as gate dielectric [31-32].

This study focuses on the formation of Mg-Silicate on AlGaN/GaN hetero-structure by annealing $MgO/SiO_2$ stacks at 850 °C, and an analysis of its physical, structural, and electrical properties. The formation of Mg-Silicate and the stoichiometry of the silicate are confirmed by XPS surface analysis. The average surface roughness of the oxide films is determined AFM analysis. The MIS diode current (I-V) and capacitance-voltage (C-V) characteristics have measured on the fabricated MIS diodes to check the electrical properties of the Mg-Silicate. The $\varepsilon_{ox}$ and interface trap density ($D_{it}$) have estimated from the C-V analysis.

## 2. Experimental Details

### 2.1. Formation of Magnesium Silicate (Mg-Silicate)

The AlGaN/GaN heterostructure, grown by Plasma Assisted Molecular Beam Epitaxy (PAMBE) technique, consists of a 2 nm GaN cap layer, a 30 nm $Al_{0.28}Ga_{0.72}N$ barrier layer, and a 0.5 nm AlN spacer layer. It also includes a 160 nm GaN buffer and a 1500 nm AlGaN transition layer grown over 4H-SiC substrate. For the formation of magnesium silicate, a six-time repeated deposit of $MgO/SiO_2$ (0.5 nm each) stack was deposited using sputtering technique, followed by annealing at 850 °C in N2 ambiance for 70 seconds. The physical characterization of both annealed and un-annealed samples was carried out using XPS, and AFM.

## 2.2. Device fabrication

The MOS diode fabrication process involved creating ohmic and Schottky contacts. Six layers of MgO/SiO$_2$ stack (0.5 nm each) were blanket deposited on the heterostructure using a sputter tool and subsequently annealed at 850 °C in a N$_2$ environment for 70 seconds to form Mg-silicate. Following this, lithography patterns were utilized to etch Mg-Silicate with BHF to establish ohmic contact. The ohmic contacts at source and drain were formed by depositing a Ti/Au/Al/Ni/Au metal stack in an electron-beam evaporator under high vacuum followed by annealing at 850 °C for 30 seconds in an N$_2$ environment [33].

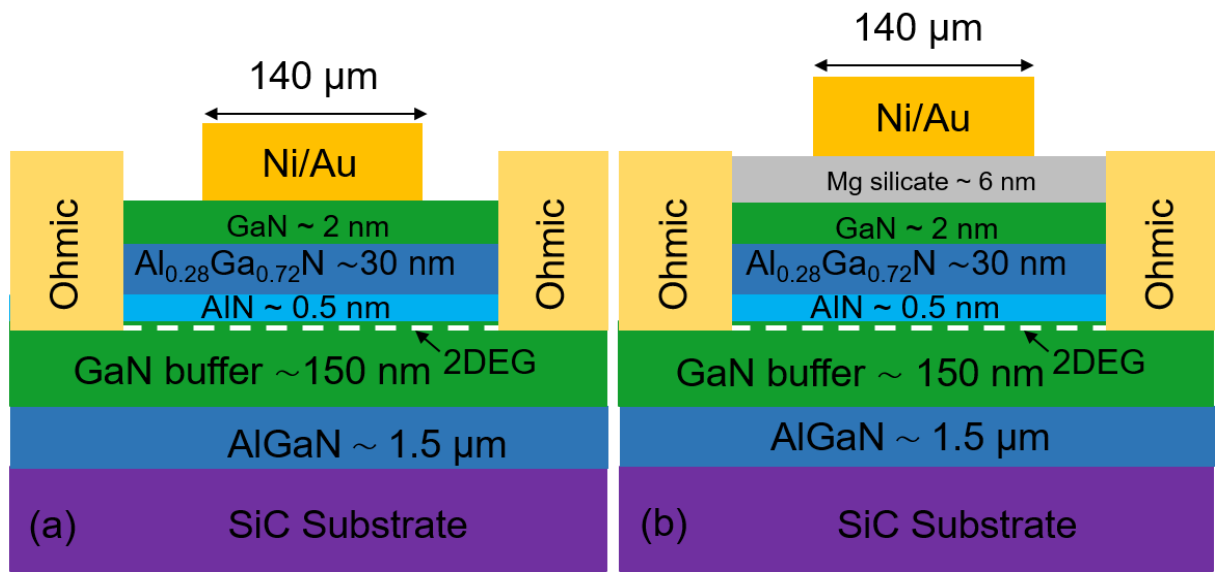

Fig. 1. Cross-sectional schematic of (a) control sample (b) MOS diode on AlGaN/GaN heterostructure.

The Ni/Au metal stack was deposited on the patterned Magnesium-silicate using an electron-beam evaporator. All the process steps are described in Fig. 2(a)-(f). The electrical characteristics were carried out using an Agilent B1500A semiconductor device analyser.

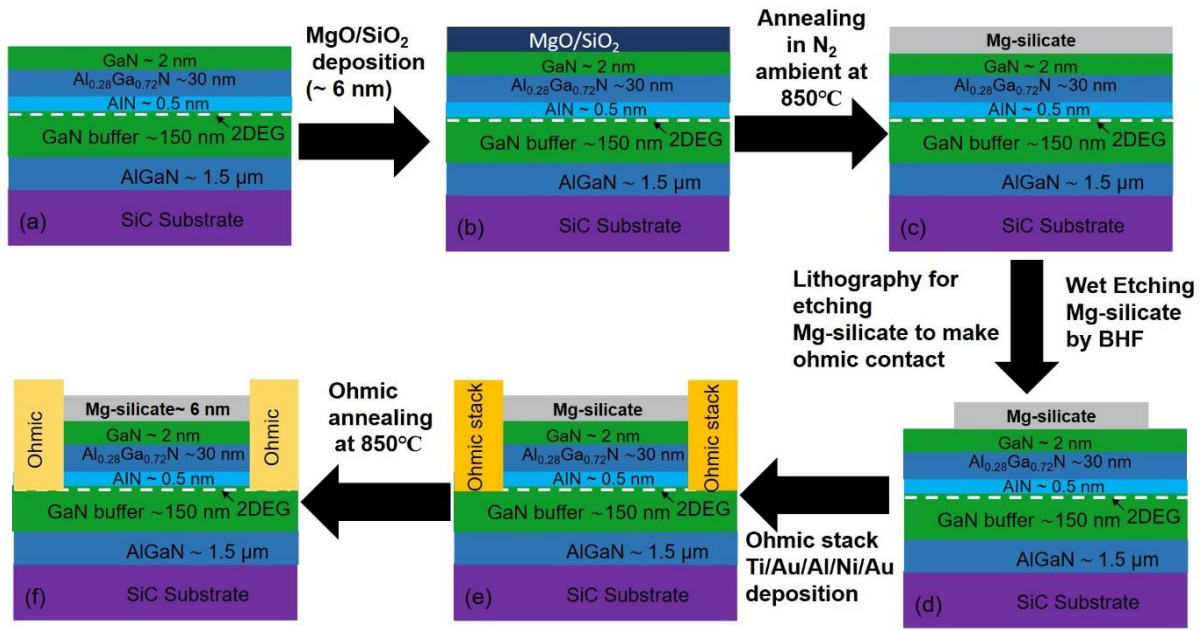

Fig. 2. (a)-(f) Process flow for the formation of Mg-Silicate in area selective regions.

## 3. Results and discussion

### 3.1 Physical characteristics of thermal oxide

The XPS analysis of Mg-silicate is shown in Fig. 3. We have taken C-1s peak at 284.8 as a reference for all the XPS data [see Figs. 3(a)–(f)].

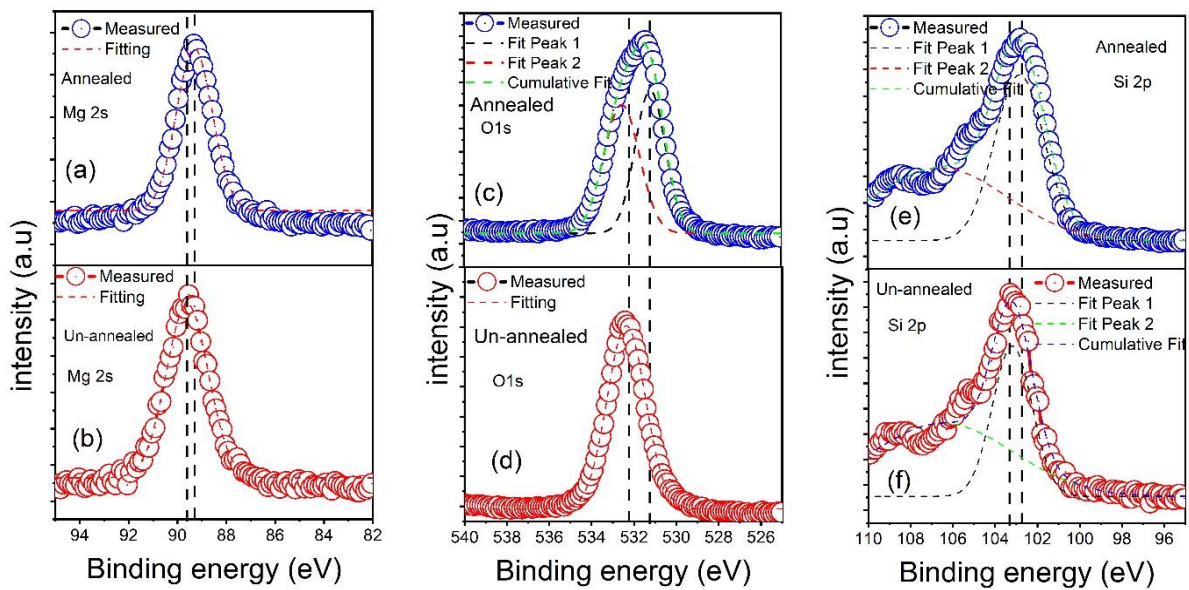

Fig. 3. (a)-(b) XPS data of Mg 2s and (c)-(d) $O_{1s}$ peaks and (e)-(f) shows Si 2p of the annealed and un-annealed samples ($MgO/SiO_2$).

Both the samples (annealed and un-annealed of $MgO/SiO_2$) show Mg 2s peaks. The Mg 2s peaks for annealed and un-annealed samples are located at 89.20 and 89.50 eV, respectively as shown in fig 3 (a)-(b). The Mg-O O1s peak for the annealed (Mg-Silicate) sample is observed at 531.43 eV, as shown in Fig. 3(c), while the Mg-O O1s peaks for the un-annealed (MgO/SiO$_2$) samples are seen at 532.40 eV, depicted in Fig. 3(d). In the SiO$_2$ structure, the Si-O bond demonstrates an O1s peak range of 532.5 - 533.4 eV, and the Mg-O bond in MgO exhibits an O1s peak at 530.0 eV. The un-annealed sample (MgO/SiO$_2$), due to its multiple stacks of MgO and SiO$_2$, displays a combined effect in the O1s peak at 532.40 eV. Meanwhile, in the annealed sample (Mg-silicate), the MgO and SiO$_2$ react at 850 °C to form Mg-Silicate, resulting in an O1s peak at 531.43 eV. The Si 2p peak range in silicates is between 102-103 eV. The Si 2p peak of the annealed sample appears at 102.60 eV, suggesting that the MgO/SiO$_2$ stack undergoes a reaction post-annealing to form Mg-silicate as indicated in Fig 3(e)-(f). The un-annealed sample, on the other hand, exhibits an Si 2p peak at 103.3 eV, attributable to the SiO$_2$ layer in the MgO/SiO$_2$ stack. The stoichiometry for the Mg-silicate is determined by integrating the corresponding peaks after normalization with atomic sensitivity factors (ASF).

The elemental atomic percentages of Mg-Silicate are calculated by normalizing the area under Mg 2s, O 1s, and Si 2p peaks accounting ASF. In Mg-silicate, the elemental atomic percentages of Mg, O, and Si are respectively 16.80%, 62.60%, and 20.60%, as outlined in Table I. The ratio of Mg/Si/O in the Mg-Silicate sample stands at 1.0/1.20/3.70. The ideal stoichiometry of Mg silicate is MgSiO$_3$, but we obtained Mg silicate stoichiometry MgSi$_{1.2}$O$_{3.7}$. The root mean square (RMS) surface roughness for both the annealed sample (Mg-silicate) and the control sample (GaN/AlGaN/GaN/SiC), as determined by AFM, is 2.2 nm and 2.7 nm respectively, as depicted in Figs. 4(a)-(b). There is a slight increase in the RMS roughness of Mg-silicate compared to the control sample, attributable to the annealing process.

**Table I**
**Elemental Atomic Percentages in Mg-silicate**

| Elements | Mg | O | Si |
| --- | --- | --- | --- |
| Percentage (%) | 16.80 | 62.60 | 20.60 |
| Ratio | 1.00 | 3.70 | 1.20 |

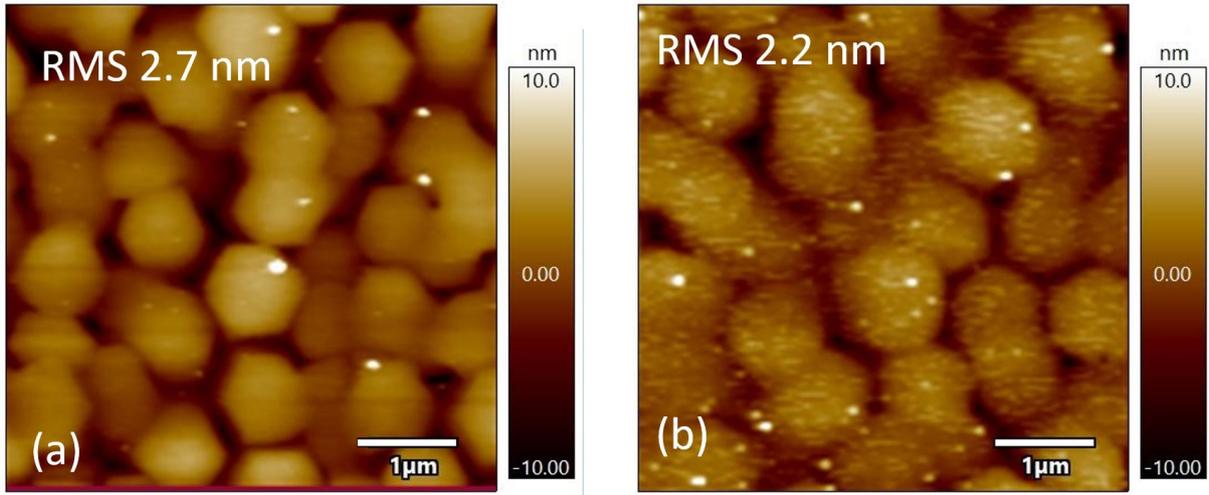

Fig. 4. AFM images of (a) control sample (GaN/AlGaN/SiC), (b) Mg-silicate sample.

## 3.2 Electrical characteristics

The capacitance-voltage (C-V) and current-voltage (I-V) measurements were performed on large area Schottky diodes (140 µm) to scrutinize the gate characteristics of the MOS-HEMTs and evaluate the quality of the silicate as a dielectric. In the MOS-Diode sample, AlGaN and Mg-Silicate capacitances are in series. Fig. 5 illustrates the C-V characteristics of both control and Mg-silicate samples at a frequency of 1 MHz The dielectric constant of Mg-silicate ($\varepsilon_{Mg-silicate}$), approximately 6.6, was determined from the inversion region capacitance using the formulas $\frac{1}{C_T} = \frac{1}{C_{control}} + \frac{1}{C_{Mg-silicate}}$ and $C_{Mg-silicate} = \frac{\varepsilon_{Mg-silicate} \times \varepsilon_0}{d} \times A$, here $C_T$, $C_{control}$, $C_{Mg-silicate}$, $\varepsilon_0$, A and d are total capacitance (0 V), control sample capacitance (0 V), permittivity of free space, area of Schottky contact of MOS-diode and thickness of the Mg-silicate respectively. The control sample capacitance ($C_{control}$) and total capacitance of Mg-silicate/AlGaN heterostructure ($C_T$) at applied bias 0 V were observed to be 318 and 240nF/cm$^2$ respectively as shown in Fig. 5.

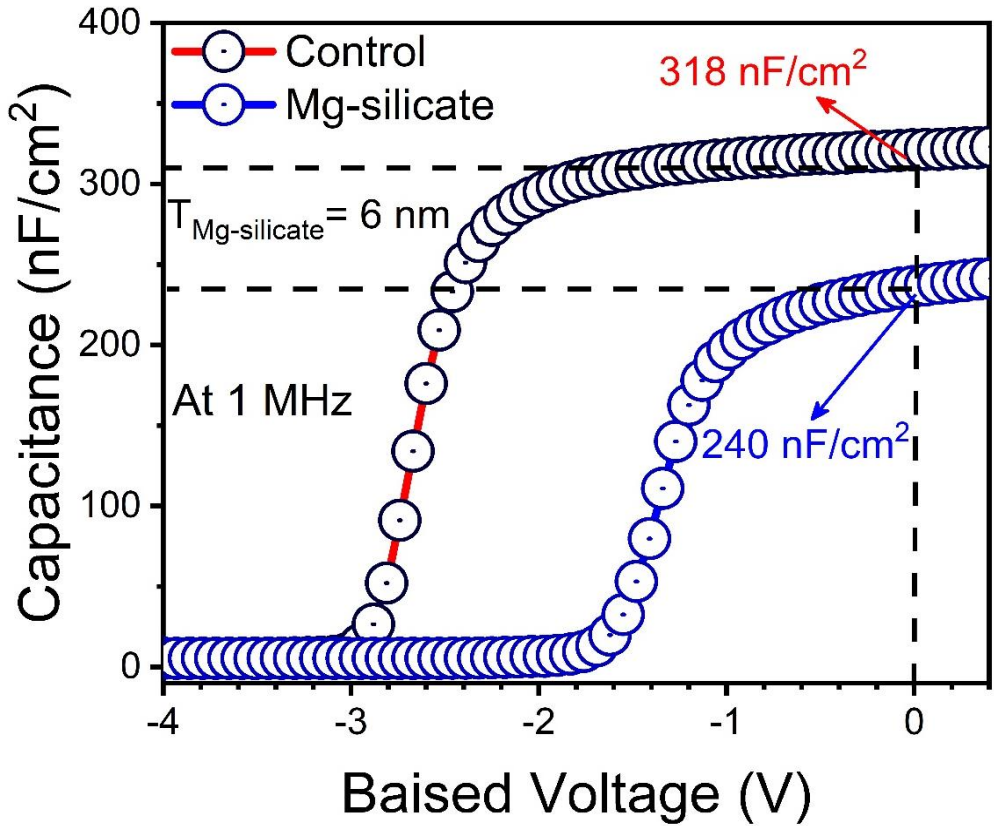

Fig.5 *C–V* characteristics at 1 MHz of control and Mg-silicate MOS diodes.

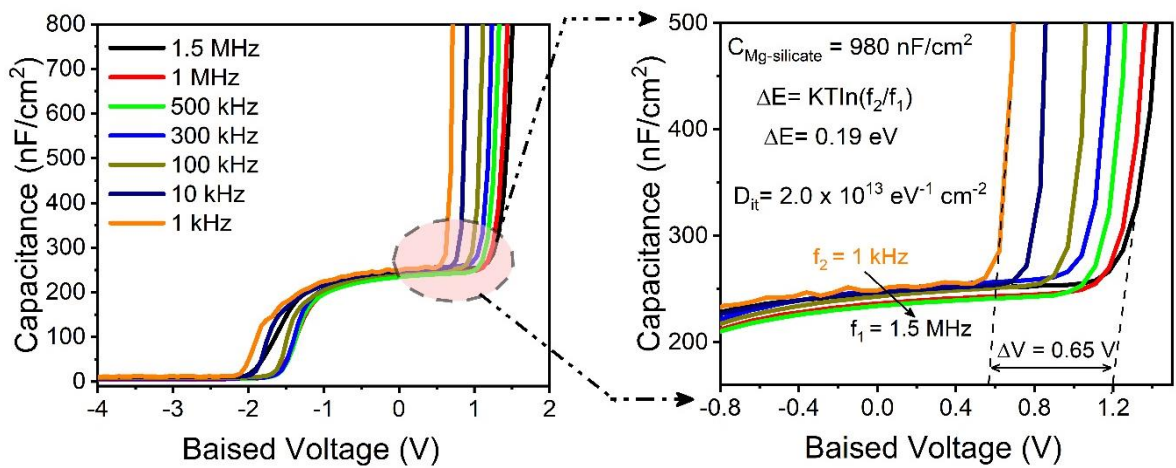

Fig. 6. (a) Frequency dependent *C–V* characteristics of Mg-silicate sample, (b) Dit calculation of Mg-silicate with AlGaN from *C-V*.

The C-V graphs for the Mg-silicate sample exhibit two sharp changes in slope when voltage is applied. The first rise relates to the formation of 2 DEG at the AlGaN/GaN interface and the second rise relates to electron accumulation at the Mg-silicate/AlGaN interface as shown in Fig 6 (a).

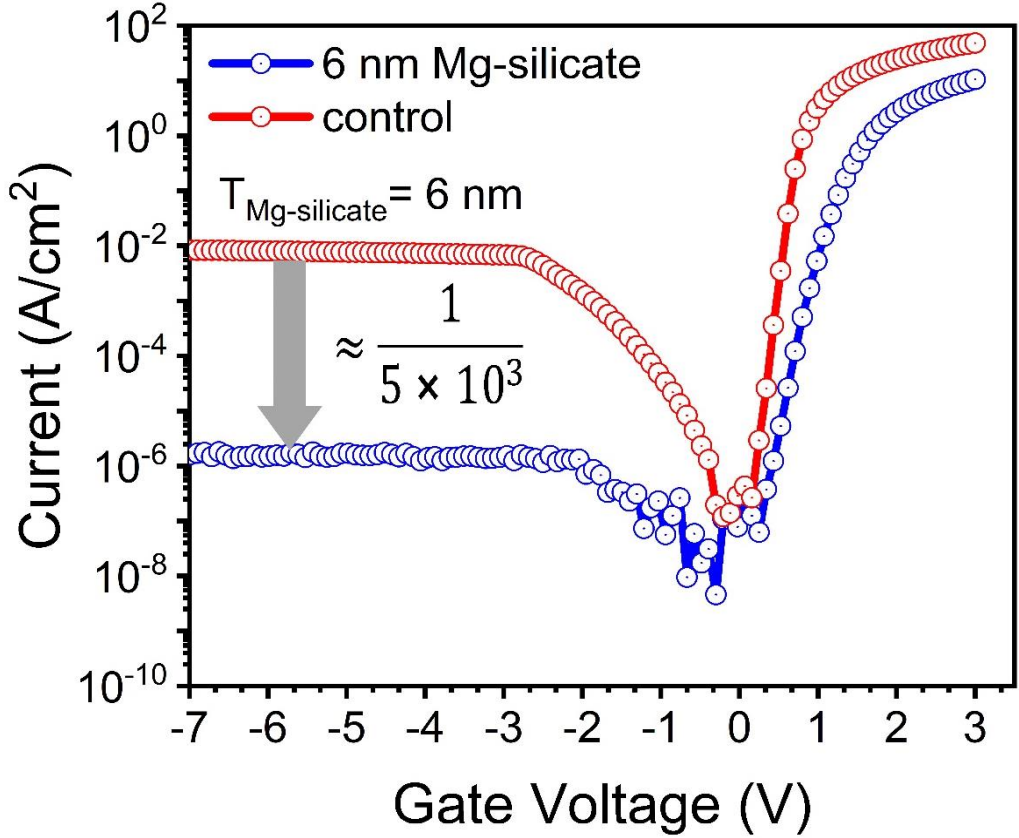

Fig. 7. I-V characteristics for the Mg-silicate and control samples.

The interface trap density ($D_{it}$) has been calculated from the C-V dispersion utilizing the equation $D_{it} = \frac{C_{Mg-silicate}}{q \times \Delta E} \times \Delta V$ [34-35], where $\Delta V$ represents the voltage difference between the frequencies ($f_1$ and $f_2$) for a fixed capacitance, corresponding to an interface charge density ($Q_{it}$) between energies $E_1$ and $E_2$. The frequency dispersion at the second rise of the C-V curve is illustrated in Fig. 6(b). The calculated $D_{it}$ for the Mg-silicate is approximately $2.0 \times 10^{13}$ cm$^{-2}$eV$^{-1}$. A frequency dispersion is also observed in the first rise of the C-V plot, potentially attributable to the charge in Mg-silicate shifting the CV curve to the right as frequency increases. Fig.7 presents the I-V characteristics of the MOS diode compared to the control sample. A decrease in reverse leakage current from $8.7 \times 10^{-3}$ A/cm$^2$ in the control sample to $1.6 \times 10^{-6}$ A/cm$^2$ in Mg-silicate-based MOS diodes at -7 V is observed, as displayed in Fig.7.

## 4. Conclusion

This study successfully demonstrates fabrication of high-quality Mg-silicate through the annealing of an MgO/SiO$_2$ stacks, characterized by a surface roughness of 2.2 nm, dielectric constant of approximately 6.6. A notable three orders of magnitude reduction in the reverse leakage current was observed in the Mg-silicate-based MOS diode, as compared to the Schottky diode (control sample). An interface trap density of approximately 2.0✕10$^{13}$ cm$^{-2}$eV$^{-1}$ was estimated. The results suggest the promising potential of Mg-silicate as an alternative gate dielectric for high-performance AlGaN/GaN MOSHEMTs.

## 5. Acknowledge

The authors would like to thank the Ministry of Electronic and Information Technology (MEITY), Govt. of India (Project: NNetRA) for financial support and IIT Bombay Nanofabrication (IITBNF) for providing fabrication and characterization facility.